\def\targetformat{arXiv}		% this flag tells it to output for arXiv, turn it off for prl style (with no appendices).
\ifx\targetformat\undefined	% seperate main text / supplemental
	\def\clsstyle{prl} 
	
	\newcommand{\appref}[1]{the Supplemental Materials}
	\newcommand{\arxivtext}[1]{}
	\newcommand{\prltext}[1]{#1}
\else		% for arXiv
	\def\clsstyle{prb} 
	
	\newcommand{\appref}[1]{App.~\ref{#1}}
	\newcommand{\arxivtext}[1]{#1}
	\newcommand{\prltext}[1]{}
	
\fi

\documentclass[aps,\clsstyle,reprint,twocolumn,showpacs,superscriptaddress]{revtex4-1}
\usepackage{graphicx,amssymb,amsmath,wasysym,psfrag,color}
\graphicspath{{img/}}
\usepackage{tikz}
\usepackage{mathtools}
\definecolor{darkgreen}{rgb}{0,0.5,0}
\newcommand{\be}{\begin{equation}}
\newcommand{\ee}{\end{equation}}

\definecolor{Zcolour}{rgb}{0.992, 0.588, 0.22}
\definecolor{dkgreen}{rgb}{0,0.5,0}
\definecolor{purple}{rgb}{0.5,0,0.5}

\usepackage[normalem]{ulem} %%temporary for edit notes

\begin{document}
	
	\title{Dynamics of the Kitaev-Heisenberg Model}
	\author{Matthias Gohlke}
	\thanks{These authors contributed equally to this work.}
	\affiliation{Max-Planck-Institut f\"ur Physik komplexer Systeme, 01187 Dresden, Germany}
	\author{Ruben Verresen}
	\thanks{These authors contributed equally to this work.}
	\affiliation{Max-Planck-Institut f\"ur Physik komplexer Systeme, 01187 Dresden, Germany}
	\affiliation{Technische Universit\"at M\"unchen, 85747 Garching, Germany}
	\author{Roderich Moessner}
	\affiliation{Max-Planck-Institut f\"ur Physik komplexer Systeme, 01187 Dresden, Germany}
	\author{Frank Pollmann}
	\affiliation{Max-Planck-Institut f\"ur Physik komplexer Systeme, 01187 Dresden, Germany}
	\affiliation{Technische Universit\"at M\"unchen, 85747 Garching, Germany}
	\date{\today}     
	
	\begin{abstract}
		We introduce a matrix-product state based method to efficiently obtain dynamical response 
		functions for two-dimensional microscopic Hamiltonians, which 
		we  apply to different phases of the Kitaev-Heisenberg model.
		We find significant broad high energy features beyond spin-wave theory even in the ordered phases proximate to spin liquids.  
		This includes the phase with  zig-zag order of the type  observed in $\alpha$-RuCl$_3$, where we find 
		high energy  features like those seen in inelastic neutron scattering experiments.
		Our results provide an example of a natural path for proximate spin liquid features to arise at high energies above 
		a conventionally ordered state, 
		as the  diffuse remnants of spin-wave bands intersect to yield a broad peak at the Brillouin zone center.  
			\end{abstract}
	\maketitle
	
	\textbf{Introduction.} %
	The interplay of strong interactions and quantum fluctuations in spin systems can give rise to new and exciting physics. 
	A prominent example are quantum spin liquids (QSL), as fascinating as they are hard to detect: they lack  local order parameters and are instead characterized in terms of emergent gauge fields. 	%
	On the experimental side, spectroscopic measurements provide particularly useful insights into 
	such systems, in particular by probing
	the fractionalised  excitations (e.g.\ deconfined spinons) accompanying the gauge field. Such measurements
	can be related to dynamical response functions, e.g. inelastic neutron scattering to the dynamical structure factor.
	On the theoretical side, determining the ground state properties of such quantum spin models is already a 
	hard problem, and it is even more challenging to understand the dynamics of local excitations.	

	Here we present a combination of the density-matrix renormalization 
	(DMRG) ground state  method and a matrix-product states (MPS) based dynamical algorithm 
	to obtain the  response functions for generic two-dimensional spin systems.
	With this we are able to access the dynamics of exotic phases that can occur in frustrated systems.
	Moreover it is also very useful for regular ordered phases where one would conventionally use large-$S$ approximations, which in some cases cannot qualitatively explain certain high energy features \cite{dallapiazza2015,banerjee2016-2}.

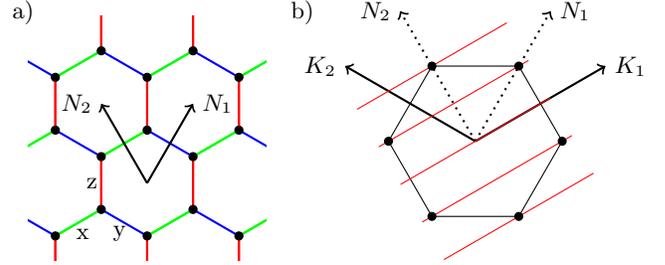
\begin{figure}
		\begin{tikzpicture}[scale=1.22]
		\begin{scope}
  			\clip(-1.3,-0.84) rectangle (1.3,1.82);
 			\foreach \x in {-2,-1, 0, 1, 2}
				\foreach \y in {-2,-1, 0, 1, 2} {
				\draw [red,thick, xshift=-\x*1cm/2+\y*1cm, yshift=\x*1.73cm/2] (0.5cm, -0.5cm/1.73)  -- (0.5cm, 0.5cm/1.73);
				\draw [red,thick, xshift=-(\x+1)*1cm/2+\y*1cm, yshift=(\x+1)*1.73cm/2] (0.5cm, -0.5cm/1.73)  -- (0.5cm, 0.5cm/1.73);
				\draw [green,thick, xshift=-\x*1cm/2+\y*1cm, yshift=\x*1.73cm/2] (0.5cm, 0.5cm/1.73)  -- (1cm, 1cm/1.73);
				\draw [green,thick, xshift=-\x*1cm/2-1cm+\y*1cm, yshift=\x*1.73cm/2] (0.5cm, 0.5cm/1.73)  -- (1cm, 1cm/1.73);
				\draw [blue,thick, xshift=-\x*1cm/2+\y*1cm, yshift=\x*1.73cm/2] (0.5cm, 0.5cm/1.73)  -- (0, 1cm/1.73);
				}
			\foreach \x in { -2,-1, 0, 1, 2}
				\foreach \y in { -2,-1, 0, 1, 2}  {	
					\filldraw [black, xshift=-\x*1cm/2+\y*1cm, yshift=\x*1.73cm/2]
								 	(0,1cm/1.73) circle (1.5pt/1.2)
									(0.5cm, 0.5cm/1.73) circle (1.5pt/1.2);
					}
		\end{scope}
		\node(xx) at (-0.7,-0.55){x};
		\node(yy) at (-0.30,-0.55){y};
		\node(zz) at (-0.6,-0.0){z};

		\draw [->,thick] (0,0) -- (-1/2,1.73/2) node [anchor=east] {$\bm ~  N_2$} ; %lattice vector a
		\draw [->,thick] (0,0) -- (1/2,1.73/2) node [anchor=west] {$\bm N_1$} ; %lattice vector b
		\node at (-1.35,1.86){a)};
	\end{tikzpicture}  \hfill
		\begin{tikzpicture}[scale=1.0]
		\draw [->,thick,dotted] (0,0) -- (-1,1.73) node [anchor=east] {$\bm N_2$} ; %lattice vector a
		\draw [->,thick,dotted] (0,0) -- (1,1.73) node [anchor=west] {$\bm N_1$}; %lattice vector b
		%\draw [->,thick,dotted] (0,0) -- ( 2,0) node [anchor=west] {$n_2$} ; %lattice vector b
		\draw [->,thick] (0,0) -- (-1.73,1) node [anchor=east] {$\bm K_2$} ; %reciprocal lattice vector 1*
		%\draw [->,thick] (0,0) -- (0,2) node [anchor=east] {$k_1$} ; %reciprocal lattice vector a*
		\draw [->,thick] (0,0) -- (1.73,1) node [anchor=west] {$\bm K_1$} ; %reciprocal lattice vector b*
		\draw (-1.1547,0) -- (-0.5773,1)--(0.5773,1) --(1.1547,0) -- (0.5773,-1)--(-0.5773,-1) -- (-1.1547,0); 
% draw cuts		
		%\draw [red] (-1.73cm/2*1.15,-1cm/2*1.15) node [anchor=east] {k-cuts}
		%	 -- (1.73cm/2*1.15,1cm/2*1.15);
		\foreach \x in {-2,-1, 0, 1, 2}
			\draw [red,xshift=-\x*sqrt(3)*1cm/6,yshift=\x*1cm/2] (-1.73cm/2*1.15,-1cm/2*1.15)  -- (1.73cm/2*1.15,1cm/2*1.15);
		
		\filldraw [black] 	(-1.1547,0) circle (1.5pt)
							(-0.5773,1) circle (1.5pt)
							(0.5773,1) circle (1.5pt)
							(1.1547,0) circle (1.5pt)
							(0.5773,-1) circle (1.5pt)
							(-0.5773,-1) circle (1.5pt)	;	% postition of dirac cones
		\node at (-2.3,1.71){b)};
	\end{tikzpicture}
	\caption{(a) Green, red and blue edges correspond to Kitaev exchange couplings $S^\gamma_i S^\gamma_j$ with $\gamma = {x,y,z}$. (b) Allowed $\mathbf k$-vectors (red lines) for an infinite long cylinder with circumference $L_2 = 6$ and periodic boundary condition along $\bm N_2$. Black nodes picture the position of the gapless Majorana cones.}
	\label{fig:BZcuts}
	\end{figure}

	We demonstrate our method by applying it to the currently much-studied 
	\emph{Kitaev-Heisenberg model} (KHM) model on the honeycomb lattice 
	\begin{equation}
	H = \sum_{\langle i,j \rangle_\gamma} K_{\gamma} S_i^\gamma S_j^\gamma +J \sum_{\langle i,j \rangle} \mathbf S_i \cdot \mathbf S_j. \label{H_KHM}
	\end{equation}
	The first term is the pure Kitaev model exhibiting strongly anisotropic spin exchange coupling \cite{kitaev2006}. 
	Neighboring spins couple depending on the direction of their bond $\gamma$ with $S^x S^x$, $S^y S^y$ 
	or $S^z S^z$ (Fig.~\ref{fig:BZcuts}).
	The second is the $SU(2)$-symmetric Heisenberg term.
	The KHM serves as a putative minimal model for several materials including Na$_2$IrO$_3$, Li$_2$IrO$_3$ \cite{chaloupka2013}, and $\alpha$-RuCl$_3$ \cite{plumb2014}. % ***[A1]
	The pure  model is an exactly solvable spin-$1/2$ model stabilizing two different Kitaev quantum spin liquids (KSL): 
	a gapped $\mathbb Z_2$ one with abelian excitations  (``A phase'') and one hosting gapless Majorana and gapped flux excitations (``B phase'') \cite{kitaev2006}. 
	If not stated otherwise, we use the parametrization $J=\cos\alpha$ and $K_{\gamma}=K=2\sin\alpha$. If $J=0$ and $K_\gamma$ bond-independent, the Kitaev model is in the B phase, which is stable under time-reversal symmetric perturbations as pointed out by Kitaev.
	Numerical studies of the ground state phase diagram of the KHM have shown an extended QSL phase for small $J$ and four symmetry broken phases for larger $J$ \cite{chaloupka2013}.

	The dynamical response functions of the pure Kitaev model are known exactly and reveal characteristic features
	\cite{knolle2014,knolle2014-1},
	such as a spectral gap due to a spin flip not only creating gapless Majorana but also gapped flux excitations.  
	This feature is perturbatively stable to small $J$ \cite{song2016}, but the influence of $J$ on high-energy features (or non-perturbatively at low energies) is unclear and of ongoing interest\cite{gotfryd2016}.
	%	 it is unclear how  high-energy features are changed by $J$, or even the low-energy features for non-perturbative $J$.
	%
	More pressingly, there appear to be proximate spin liquids \cite{suzuki2015,banerjee2016}, such as possibly the currently much-studied 
	$\alpha$-RuCl$_3$ \cite{plumb2014,johnson2015,kim2016,sandilands2016,sandilands2016-1,chaloupka2016,lang2016,halasz2016,koitzsch2016,banerjee2016,banerjee2016-2}, %***[A1-A9] 
	whose low-energy physics is consistent with spin waves on an ordered background, but whose broad high-energy features resemble those of a KSL. 
	In particular, for intermediate energy scales there are star-like features \cite{banerjee2016-2} apparently arising from a combination of spin wave and QSL physics. 
	%
	%Previous work on the dynamics of such Kitaev-like models has used exact diagonalization \cite{suzuki2015,gotfryd2016}.
	
	In this article, we first revisit the ground state phase diagram and confirm the previously found phases.
	The infinite cylinder geometry allows us to numerically confirm that the 
		gaplessness  of the KSL is robust throughout the entire phase.
	Secondly we use a recently introduced MPS based time evolution algorithm \cite{zaletel2015} to obtain the dynamical spin structure factor.
	We benchmark our method by comparing to exact results for the Kitaev model and find a good agreement.
	We calculate the spectra of different (non-soluble) phases of the KHM.
	Most notably, we identify broad high energy continua even in ordered phases that are reminiscent of the broad features 
	observed in recent experiments on $\alpha$-RuCl$_3$ and which are moreover similar to the high energy features in the spin 
	liquid phase, thus providing a concrete realisation of the concept of a proximate spin liquid. 
		
	%%%%%%%%%%%%%%%%%%%%%%%%%%%%%%%%%%%%%%%%%%%%%%%%%
	%%%%%%%%%%%%%%% Ground state %%%%%%%%%%%%%%%%%%%%%%%%%%%%
	%%%%%%%%%%%%%%%%%%%%%%%%%%%%%%%%%%%%%%%%%%%%%%%%%
	
	\begin{figure}[tb]
		\includegraphics[width=8.5cm]{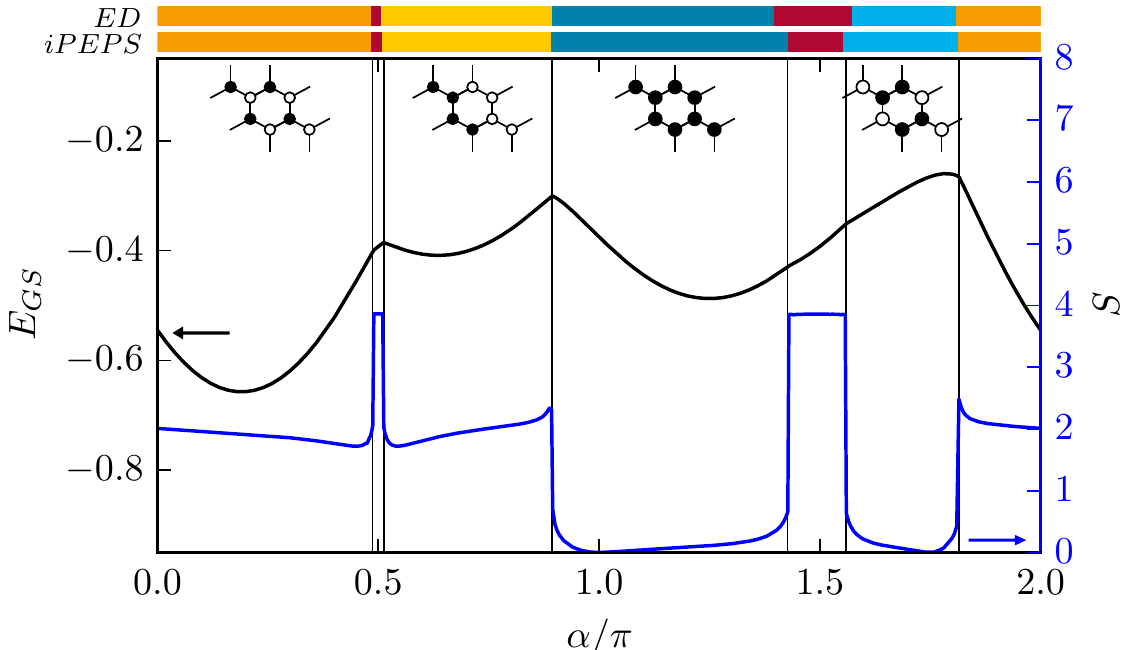}
		\caption{Phase diagram for an infinite cylinder with circumference $L_2 = 12$ obtained using iDMRG. The black line corresponds to the ground state energy density and the blue line to the entanglement entropy for a bipartition of the cylinder into a left and right half. The insets illustrate the ordering pattern of the magnetic phases. Two spin liquid phases exist around the pure Kitaev model ($\alpha = 0.5 \pi$ and $1.5\pi$). The results of ED \cite{chaloupka2013} and iPEPS \cite{osorioiregui2014} are illustrated on top.}
		\label{fig:phasediagram}
	\end{figure}
	\textbf{Ground state phase diagram.}
	We use the iDMRG algorithm on the KHM on infinite cylinders to map out the phase diagram. 
	We choose cylinder geometries such that the corresponding momentum cuts contain the gapless Majorana modes of the Kitaev spin liquid. 
	For the pure isotropic Kitaev model, there are gapless Majorana cones on the corners of the first Brilluoin zone, Fig.~\ref{fig:BZcuts}b. 
	The full KHM has a $C_6$ symmetry which means that in the 2D limit these cones cannot shift.
	The iDMRG method determines the ground state of systems of size $L_1 \times L_2$ where $L_1$ is in the thermodynamic limit and  $L_2$ a finite circumference of up to $12$ sites beyond what is achievable in exact diagonalization.
	While traditionally iDMRG is used for finding the ground state of one-dimensional systems, 
	it has become  a fairly unbiased method for studying two-dimensional frustrated systems.
	The resulting phase diagram for  $L_2 = 12$ is shown in Fig.~\ref{fig:phasediagram} (for the iDMRG simulations we keep $\chi = 1200$ states), which agrees with previous studies \cite{chaloupka2010,jiang2011,chaloupka2013,sela2014,osorioiregui2014,shinjo2015}.
	For this $L_2$, the system is compatible with the sublattice transformation that maps zigzag to AF and stripy to FM\cite{chaloupka2010}.
	Plotted are the ground state energy and the entanglement or von-Neumann entropy $S=-\mathrm{Tr} \rho^{\mathrm{red}} \log \rho^{\mathrm{red}}$ of the reduced density matrix $\rho^{\mathrm{red}}$ for a bipartitioning of the cylinder by cutting along a ring. 
	Both the cusps in the energy density and the discontinuities of the entanglement entropy indicate first order transitions. 
	A careful finite size scaling is difficult because of the large bond dimension needed and thus it is not possible to make definite statements about whether the transitions remain first order in the limit $L_2\rightarrow\infty$.   
	The symmetry broken phases can  be identified by measuring the local magnetization.
	We identify a N\'{e}el phase ($-0.185 < \alpha/\pi < 0.487$) that extends around the pure anti-ferromagnetic Heisenberg
	 \footnote{Note that due to the mapping of the 2D lattice onto a 1D chain, Mermin-Wagner-Coleman\cite{coleman1973} applies at the pure AF-Heisenberg point and suppresses long range N\'{e}el order. In fact it is replaced by a gapped symmetry-preserving state, which extends over a finite region in parameter space. For further discussion,
	\ifx\targetformat\undefined	% seperate main text / supplemental
	  	see supplemental material. % \cite{supplement}.} 
	\else
		see appendix \ref{app:1Dvs2D}.
	\fi	
	}
	 point, the corresponding zigzag phase  ($ 0.513 < \alpha/\pi < 0.894 $), a ferromagnetic phase around the pure FM Heisenberg point ($ 0.894 < \alpha/\pi < 1.427 $), and its  stripy phase ($1.559 < \alpha/\pi < 1.815$).
	The two KSLs between N\'{e}el and zigzag as well as between FM and stripy are confirmed to be gapless. %
	In particular, if $L_2$ is a multiple of six we use the finite entanglement scaling approach \cite{calabrese2004,tagliacozzo2008,pollmann2009} and extract the expected chiral central charge $c=1$ for both KSLs%
	\ifx\targetformat\undefined	% seperate main text / supplemental
		\cite{supplement}, each of the two Majorana cones contributing $c=1/2$.
	\else
		, each of the two Majorana cones contributing $c=1/2$.
		See also appendix \ref{app:entanglement_scaling}.
	\fi
	Note that when a gapless spin liquid is placed on a cylinder, the gauge field  generically adjusts to open  a gap \cite{he2016}.
	In order to see gapless behaviour, we have to initiate the iDMRG simulations in the gapless sector to access a metastable state 
	\ifx\targetformat\undefined	% seperate main text / supplemental
		\cite{supplement}.
	\else
		(see appendix \ref{app:ksl_sectors} for additional details).
	\fi	
	The gapped ground state having a non-zero flux through the cylinder overestimates the stability of the QSL phases. 
	%\ruben{Shall we explicitly mention that we would get different/wrong phase transition values if we worked within the gapped sector?} \matthias{Do you agree with my two sentences?}
	%
	It is notable how well the phase boundaries agree with those from the infinite projected entangled pair state (iPEPS) simulations \cite{osorioiregui2014}. 
	
	%%%%%%%%%%%%%%%%%%%%%%%%%%%%%%%%%%%%%%%%%%%%%%%%%%
	%%%%%%%%%%%%%%% Dynamics %%%%%%%%%%%%%%%%%%%%%%%%%
	%%%%%%%%%%%%%%%%%%%%%%%%%%%%%%%%%%%%%%%%%%%%%%%%%%
	
	\begin{figure}
		\begin{tikzpicture}
			\node at (0,0) [anchor=north west]  {\includegraphics[scale=.4,trim=0cm 1.8cm 0cm 0.3cm,clip]{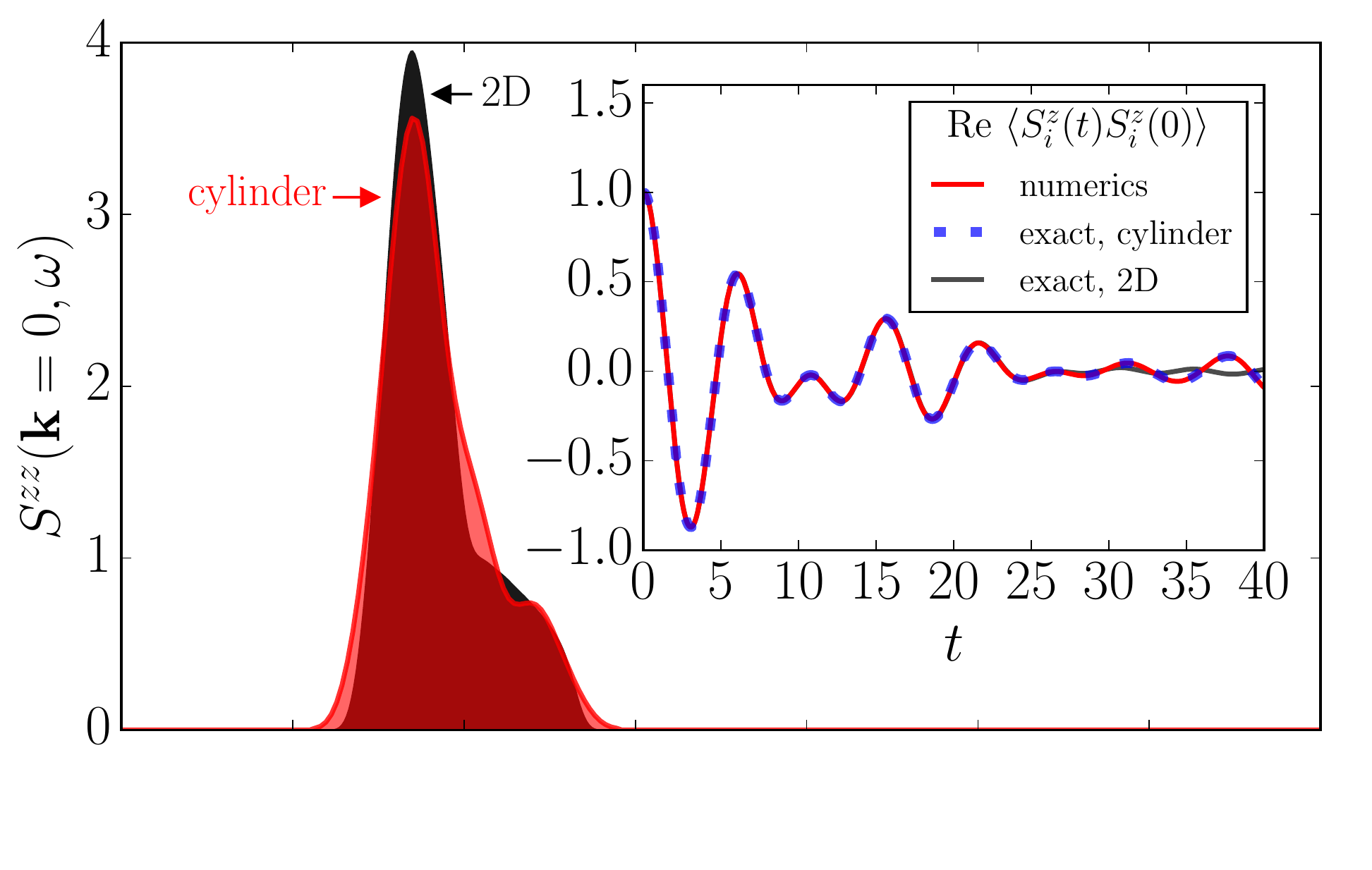}};
			\node at (0,0) [anchor=north west] {a)};
		\end{tikzpicture}
		\begin{tikzpicture}
			\node at (0,0) [anchor=north west]  {\includegraphics[scale = .4,trim=0cm 0.5cm 0cm 0.3cm]{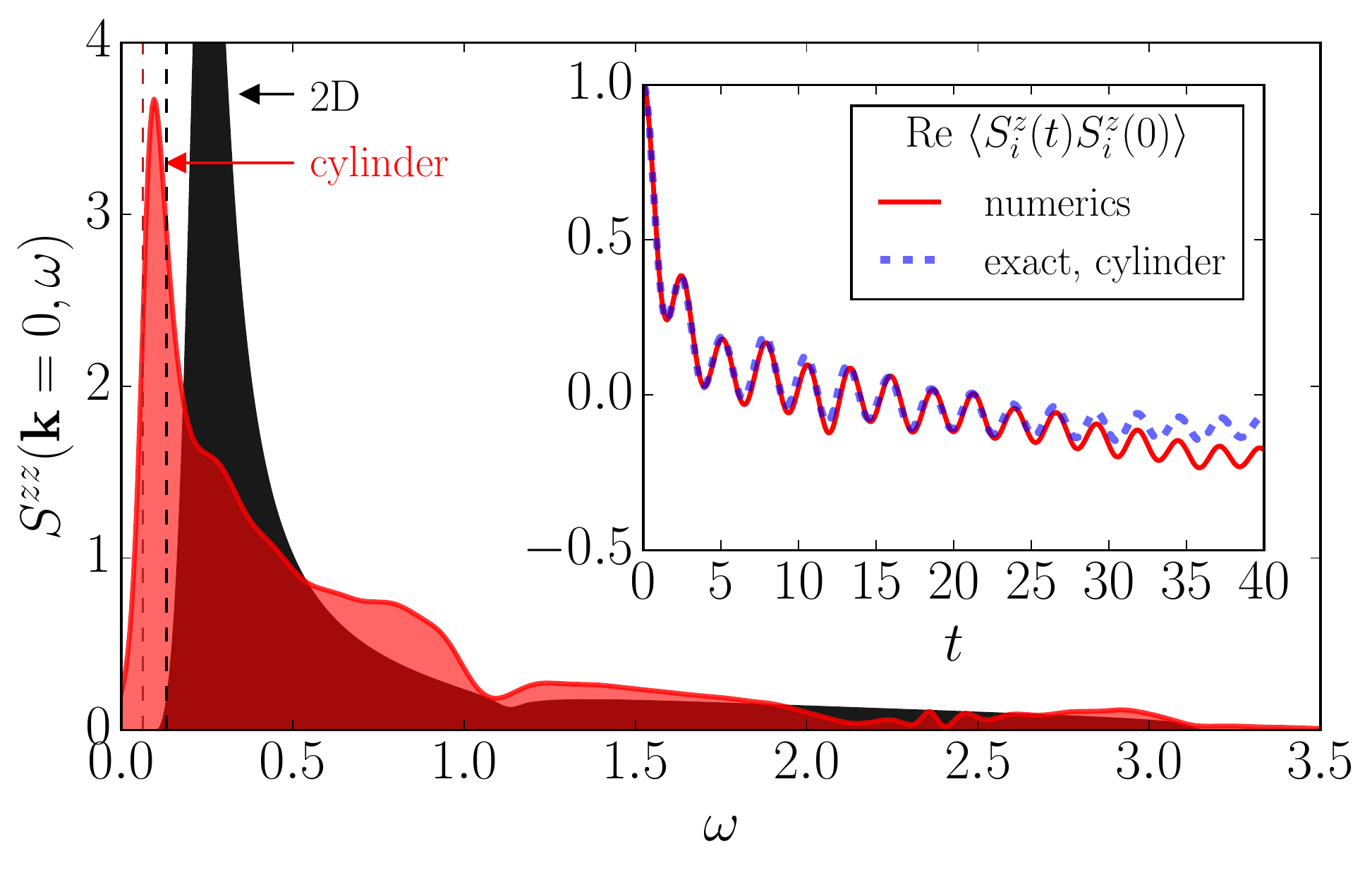}};
			\node at (0,0) [anchor=north west] {b)};
		\end{tikzpicture}
		\caption{
		Dynamical structure factor $\mathcal S^{zz}(\boldsymbol k=0, \omega)$ from our numerical approach compared with exact result (insets show real time data). Exact results were obtained following [\onlinecite{knolle2014}], except for the blue curve in (b)[\onlinecite{knolle_unpub}]. % figure \ref{fig:exact_KSL}(b) \cite{knolle_unpub}.
		(a) Gapped KSL on a cylinder with $L_2 = 10$ and anisotropic couplings $K_x = -2$ and $K_y = K_z = -\frac{1}{3}$.
		(b) Gapless isotropic KSL with $L_2 = 6$ and $\alpha = \frac{3\pi}{2}$.
	%Dynamical structure factor $\mathcal S(\boldsymbol k, \omega)$ obtained numerically with a MPO based time evolution of
	%(a) a gapped KSL on a cylinder with $L_2 = 10$ and anisotropic couplings $K_x = -2$ and $K_y = K_z = -\frac{1}{3}$, and
	%(b) a gapless isotropic KSL with $L_2 = 6$ and $\alpha = \frac{3\pi}{2}$.
	%Additionally, exact results in the 2D limit\cite{knolle2014} are plotted.
	%The insets show the corresponding real time data obtained numerically, exact in the 2D limit and exact on a cylinder\cite{knolle_unpub}.
		}
		\label{fig:exact_KSL}
	\end{figure}      
	
	\textbf{Dynamical structure factor $\mathcal{S}(\boldsymbol k,\omega)$.} 
	Starting from a ground state obtained using  iDMRG, we  calculate $\mathcal{S}(\boldsymbol k,\omega)$ 
	by Fourier transforming the dynamical correlation 
	function $C^{\gamma\gamma}(\mathbf r, t) = \langle S^\gamma_{\mathbf r} (t) S^\gamma_{\mathbf 0}(0) \rangle.$
	The real-time correlations can be efficiently obtained using a recently introduced matrix-product operator based time evolution method \cite{zaletel2015}.
	This allows for long range interactions resulting from unraveling the cylinder to a one-dimensional system which render standard methods like the time-evolving block decimation inefficient.
	Following the general strategy laid out in Refs.~[\onlinecite{kjall2011,phien2012,zauner2015}], we perform 
	the simulations for an infinite cylinder with a fixed circumference.
	Note that the entanglement growth and the resulting growth of the required number of states is generically slow as we 
	only locally perturb the ground state and thus long times can be reached even in the cylinder geometry.
	We show results obtained for $0\leq t \leq T$
	and to avoid  Gibbs oscillations we multiply our real-time data with a Gaussian ($\sigma_t \approx 0.43 T$). This 
	corresponds to a broadening in $\omega$-space ($\sigma_\omega \approx \frac{2.3}{T}$). 
	We use linear prediction to allow room for 
	the tail of the Gaussian in real-time, but confirm that the final results do not depend on its details \cite{white2008}.
	Thence, \[\mathcal S^{\gamma \gamma}(\boldsymbol k,\omega) = \frac{1}{2\pi} \sum_{\mathbf r} \int_{-\infty}^\infty  e^{i(\omega t-\mathbf k \cdot \mathbf r)} C^{\gamma\gamma}(\mathbf r, t) \; \mathrm dt\]
	normalized as $\int \mathcal S^{\gamma \gamma}(\boldsymbol k,\omega) \; \mathrm d \mathbf k \mathrm d \omega = \int \mathrm d \mathbf k$. If not stated otherwise, we present results for $\mathcal S(\boldsymbol k,\omega) = \sum_\gamma \mathcal S^{\gamma \gamma}(\boldsymbol k,\omega)$.

	We benchmark the method by comparing our numerical approach to exact results for the pure Kitaev model.
	Figure~\ref{fig:exact_KSL}a shows a comparison for the gapped Kitaev model in the A phase with $K_x/K_{y,z} = 6$, the exact solution for $\mathcal S^{zz}(\boldsymbol k=0,\omega)$ shown in black. Our numerics (with resolution $\sigma_\omega \approx 0.06$ in units shown) for an infinite cylinder with $L_2= 10$ (red) agrees well with such features as gap, bandwidth and total spectral weight. In the real-time data (inset), whilst the numerics agrees with the exact solution for the cylinder geometry, it overlaps with the 2D result only until a characteristic time scale  corresponding to the perturbation traveling around the cylinder and then feeling the static fluxes inserted by the spin-flip. More generally we expect such timescales (after which 2D physics becomes 1D) to be particularly significant for systems with fractionalization.
	% info to self: so it seems v = 0.8 sites per time unit (in my numerics units; conversion factor is 2/4.5, giving v = 0.36) for slower moving part of lightcone, but this still includes an x bonder, i.e. is for on average (y/z-bond) followed by (x bond). %
	%
	For Fig.~\ref{fig:exact_KSL}b we take $K_x = K_y = K_z = -2$ being in the gapless KSL phase at $\alpha = \frac{3\pi}{2}$. Comparing the exact 2D result (black) to our numerics for a cylinder of circumference $L_2 = 6$ (red), we see qualitative similarities, such as a spectral gap (dashed lines; slightly obscured by our finite-time window), a dip where the fluxes suppress the van Hove singularity of the Majorana spectrum\cite{knolle2014},  comparable bandwidth and strong low-energy weight. To better resolve the spectral gap, we rely slightly on linear prediction \cite{white2008} by using a real-time Gaussian envelope with $\sigma_t = 0.56T$, corresponding to 
	$\sigma_\omega \approx 0.045$. Two striking quantitative differences are (i) the spectral gap which for this circumference is approximately half that of the 2D limit, and (ii) the presence of a delta-peak on this gap ($\approx 4\%$ of total spectral weight). The latter, present for any cylinder, vanishes as $L_2 \to \infty$. The inset compares  exact real-time results on the cylinder \cite{knolle_unpub} with our numerics. Despite the true ground state on this cylinder being gapless and MPS only being able to capture gapped ground states exactly, we still find  good agreement for appreciable times.
	
	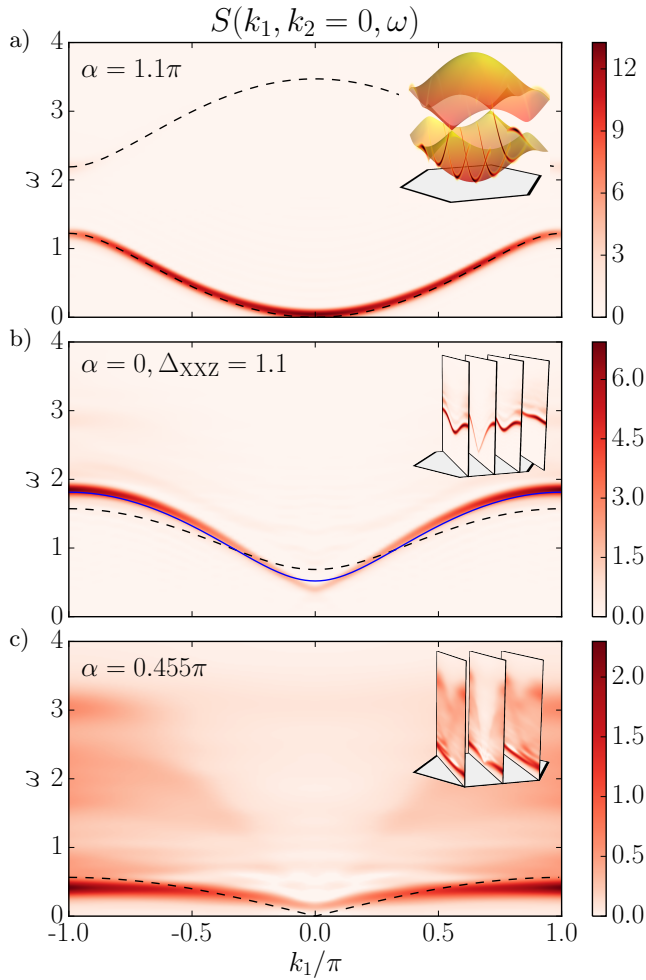
\begin{figure}
		\begin{tikzpicture}
			\node at (0,0) [anchor=north west]  {\includegraphics[scale = .52]{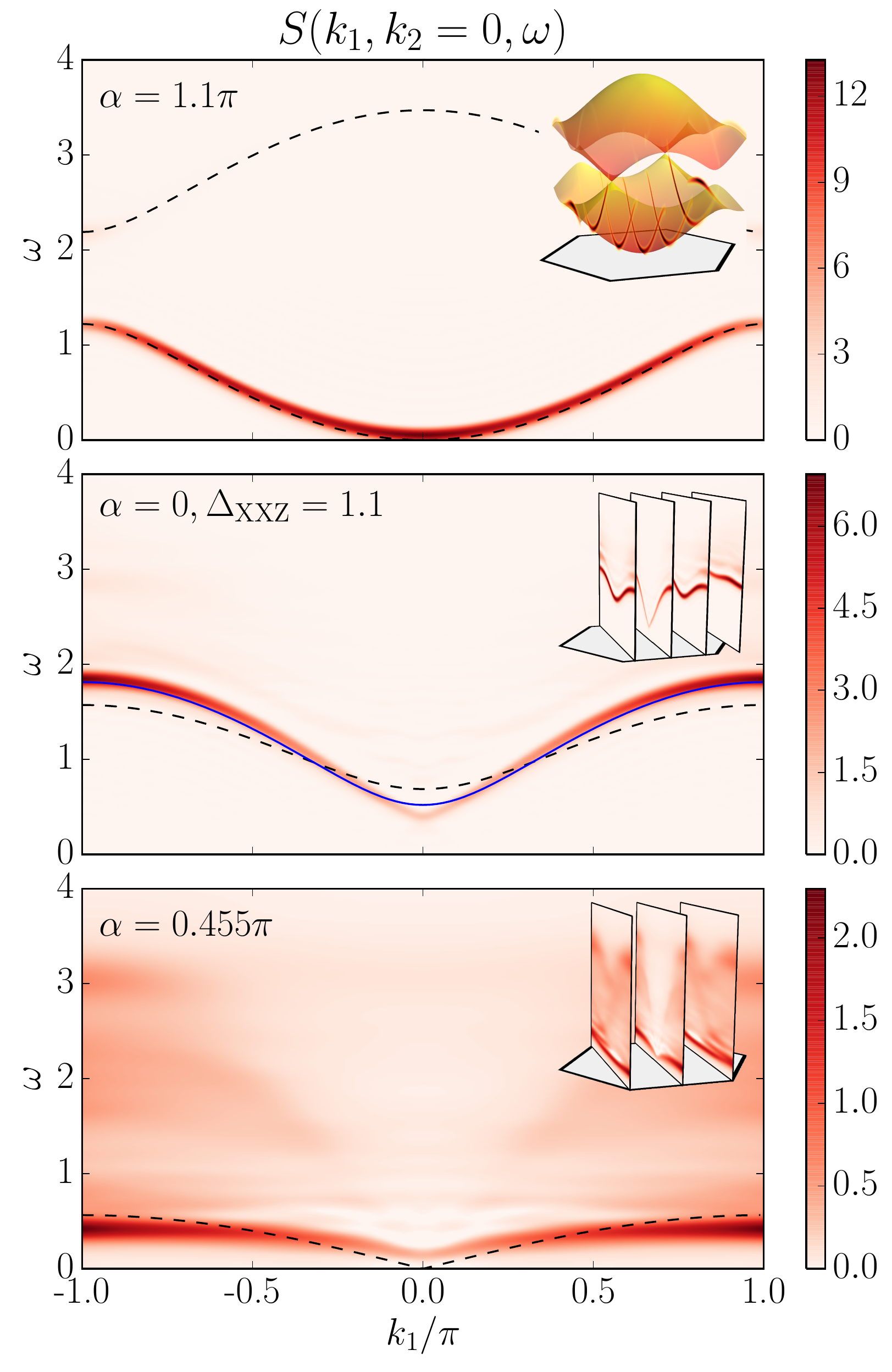}};
			\node at (0,-4mm) [anchor=north west] {a)};
			\node at (0,-43.8mm) [anchor=north west] {b)};
			\node at (0,-83.7mm) [anchor=north west] {c)};
		\end{tikzpicture}
		\caption{Dynamical structure factor  $\mathcal S(\boldsymbol k, \omega)$ for cuts $\boldsymbol k=(k_x,0)$ in different phases of the KHM with the $\omega$-resolution $\sigma_\omega \approx 0.06$. Dashed lines show results from LSWT.  Insets show the data for all allowed cuts. (a) Ferromagnetic phase for a cylinder with $L_2=12$. (b) Antiferromagnet with small spin anisotropy without Kitaev term ($L_2=8$). Blue line shows second order spin wave calculations. (c) Antiferromagnetic phase in proximity of the KSL ($L_2 = 6$).}
		\label{fig:SKHM}
	\end{figure}
	\begin{figure}[tb]
		\includegraphics[scale =.49]{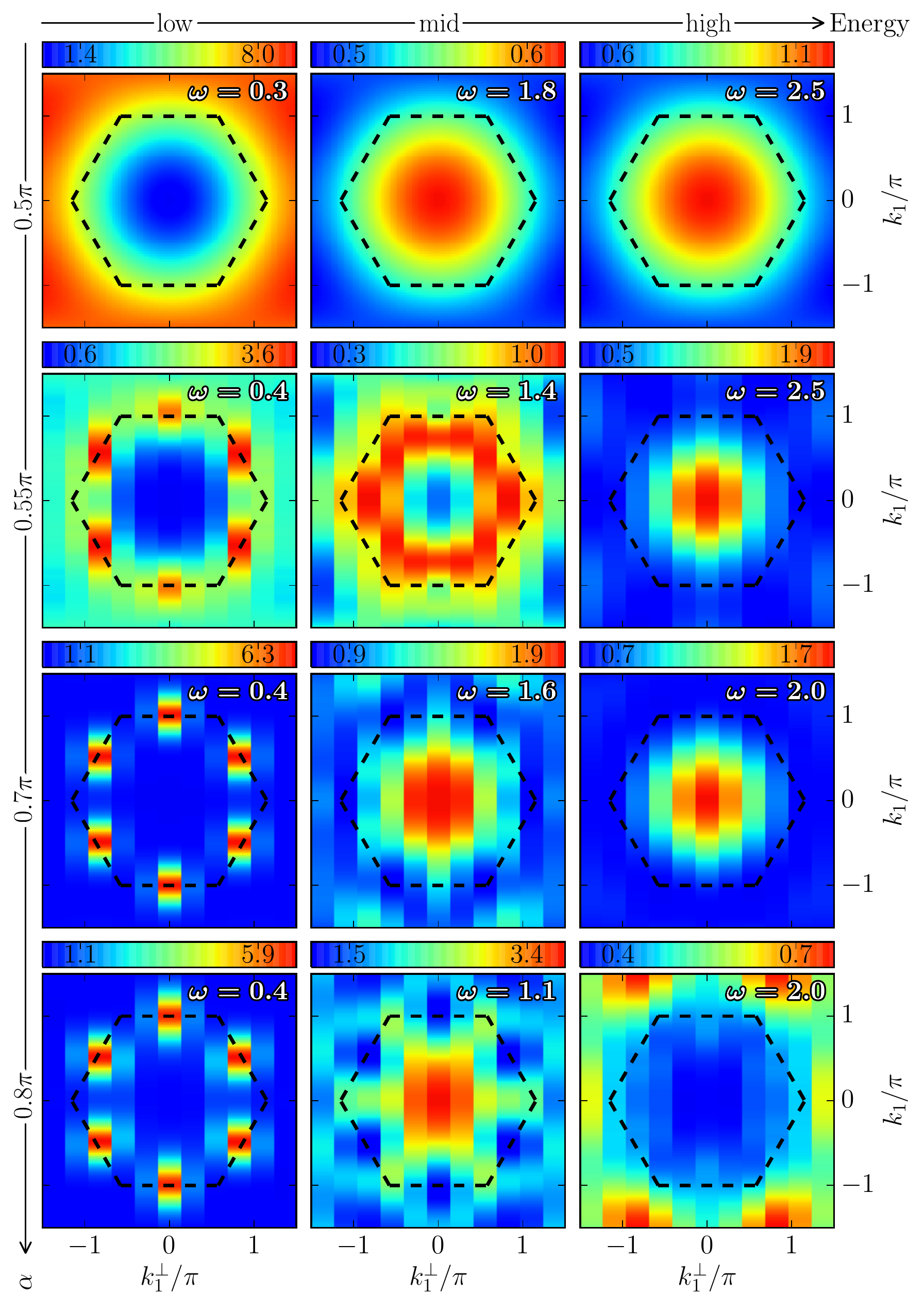}
		\caption{$S(\mathbf k, \omega)$ at three different energies for four models: KSL at $\alpha = 0.5\pi$ (analytic result, 2D) and zigzag order at $\alpha = 0.55\pi,0.7\pi,0.8\pi$ (with $L_2 = 12$)}
		\label{fig:colorful}
	\end{figure}
	After this benchmarking, we explore $\mathcal S(\boldsymbol k,\omega)$  in different phases of the KHM shown in Fig.~\ref{fig:SKHM}, all with $\sigma_\omega \approx 0.06$.
	The pure Heisenberg FM ($\alpha = \pi$) can be solved in terms of linear spin wave theory (LSWT) and numerically captured with bond dimension $\chi = 2$. Instead of this special point, in Fig.~\ref{fig:SKHM}a we show results for $\alpha = 1.1\pi$ (corresponding to $K = 0.65 J$) where we still find excellent agreement with LSWT. 
	Note that there is an extremely small gap ($\approx 0.05 |J|$) despite the presence of anisotropic couplings, as the entire KHM is $SU(2)$-symmetric in LSWT. 
	We do not observe any strong cylinder effects on the dynamics, which is presumably related to the short correlation length and the absence of fractional excitations.
	The pure Heisenberg AFM (with small XXZ anisotropy) in Fig.~\ref{fig:SKHM}b shows appreciable deviations from LSWT, with second order SWT \cite{weihong1991} giving better agreement. 
	Moreover, the weight in the spin waves is approximately halved, indicating the importance of higher order magnon contributions. 
	Staying within the N\'{e}el phase but approaching the QSL,  spin wave theory cannot even qualitatively describe Fig.~\ref{fig:SKHM}c, with much weight in very broad high energy features unaccounted for. 

	Lastly we focus on a parameter regime producing zig-zag ordering like that found 
	in $\alpha$-RuCl$_3$ \cite{johnson2015,banerjee2016, banerjee2016-2}. %***[A2]
	Fig.~\ref{fig:colorful}  shows $\mathcal S(\boldsymbol k,\omega)$ for four different choices of $\alpha$: the first row contains the exact solution for the pure AFM Kitaev model, and the subsequent rows are all numerical results within the zigzag phase with increasing $\alpha$. 
	For each $\alpha$ we show $\mathcal S(\boldsymbol k,\omega)$ at fixed $\omega$: the columns display representative low-, mid- and high-energy features, with parameters
	 $L_2 = 12$ and time cut-off $T=10$ corresponding to $\sigma_\omega \approx 0.23$. We average over the different symmetry broken directions. 
	\ifx\targetformat\undefined	% seperate main text / supplemental
		Results for $L_2 = 6$ and $T=40$ reveal that even at this resolution the high-energy features stay very broad \cite{supplement}.
	\else
		In appendix \ref{app:L2_6_T40}, we show results for $L_2 = 6$ and $T=40$, revealing that even at this resolution the high-energy features stay very broad.
	\fi	
	The first column shows the low-energy physics of the Kitaev model being reconstructed into spin wave bands, 
	with minima on the edges of the first Brillouin zone.
	For $\alpha = 0.7\pi, 0.8\pi$ these obey the $C_6$-symmetry, indicating that the cylinder geometry locally looks like 2D.
	Interestingly, the high-energy physics of the ordered phases is very similar to that of the pure Kitaev model: we have broad features centered around $\boldsymbol k = 0$ which are diffuse w.r.t. $\omega$, with its characteristic energy and width simultaneously decreasing as $\alpha$ increases. 
	The interplay between these low- and high- energy features then gives rise to different mid-energy shapes. In fact the six spin wave bands start on the edges of the first Brillouin zone. As the energy increases, these bands become increasingly diffuse, eventually overlapping in a very broad blob above the symmetric $\Gamma$ point $\boldsymbol k = 0$. Both  spin waves and  blob  sharpen as one moves away from the nearby QSL. 
Comparing with inelastic neutron data for $\alpha$-RuCl$_3$ \cite{banerjee2016-2}, we find the best qualitative agreement in Fig.~\ref{fig:colorful} around $\alpha = 0.7\pi$. 
	In particular at intermediate energies there is a six-pointed-star whose arms point towards the edges of the first Brillouin zone.
	It is interesting to note that if we do not average over different symmetry broken directions, the low-energy physics strongly breaks the $C_6$ symmetry yet the six-pointed star at intermediate energies persists: thus even if we interpret these high energy features as the overlap of broad spin waves, at this point the effect of symmetry breaking has disappeared. Under what conditions such a symmetry restoration occurs more generally is an interesting question. 

	\textbf{Conlusion.} %
	We have presented a new method for obtaining the dynamical properties of generic lattice spin models 
	in (quasi-)two dimensions, which we expect to be useful for many future studies. In the KHM, our study   
	reveals several features beyond spin-wave theory even in the ordered phases, providing a more detailed 
	picture for the concept of a proximate spin liquid as potentially realised in $\alpha$-RuCl$_3$.
	
	\textbf{Acknowledgements.} 
	We are grateful to Roser Valenti, Mike Zaletel and Johannes Knolle for stimulating discussions. In particular we thank Johannes for providing unpublished data for the dynamical correlations of the isotropic Kitaev model on the cylinder. 
	This work was supported in part by DFG via SFB 1143 and Research Unit FOR 1807 through grants no. PO 1370/2-1.
	
	\bibliography{khdyn_0108.bib}
	
	\ifx\targetformat\undefined	% seperate main text / supplemental
		% do nothing
	\else
		\pagebreak
		\appendix
		\section{1D vs 2D physics: symmetry breaking} \label{app:1Dvs2D}
	From Monte-Carlo studies \cite{reger1989} it is known that the ground state of the Heisenberg antiferromagnet (AFM) on the  honeycomb lattice displays symmetry breaking N\'eel order.
	However, when we place the Heisenberg AFM on an infinitely long cylinder of finite circumference, it is in principle a 1D system and the presence of a continuous symmetry in fact {forbids} spontaneous symmetry breaking \cite{coleman1973}.
	Instead we numerically find a gapped state which preserves both spin rotation and translation symmetry. This is analogous to the results for stacking an even number of coupled spin-$\frac{1}{2}$ Heisenberg chains \cite{white1994}.
	The transition from 1D to 2D can be understood by noting that this symmetry-preserving state is effectively N\'eel-like within a correlation length $\xi$, the latter growing with circumference. Similarly to how one determines spontaneous symmetry breaking from finite size scaling in the context of exact diagonalization, one can conclude that the 2D limit achieves N\'eel order by scaling with respect to circumference.

	The presence of a gap implies this symmetry-preserving state is stable under $SU(2)$-breaking perturbations. For example for $L_2 = 6$ it extends over $-0.2 \pi \leq \alpha \leq 0.43 \pi$, with a N\'eel order arising for larger $\alpha$ until we hit the spin liquid. The stability of this symmetry-preserving state under Kitaev perturbations is presumably related to the fact that the N\'eel order which arises in the 2D limit would have a very small spin gap. This is different for XXZ-type perturbations, which induce N\'eel order for relatively small anisotropies as shown in Fig.~\ref{fig:RVB} (with $\Delta = 1.1$), where our state is numerically converged (for large $\chi$) and the physics quickly becomes independent of circumference.

	The DMRG simulations use a parameter $\chi$ which gives an upper bound on the entanglement. By limiting $\chi$ we can find a variational state with $\xi < L_2$. Locally this state then looks 2D and hence we can have symmetry breaking even for the $SU(2)$-symmetric Heisenberg model, as confirmed in Fig.~\ref{fig:RVB}. As we increase $\chi$, eventually $\xi$ becomes of the order of $L_2$, which signals the transition of 2D to 1D physics and the symmetry-preserving state arises. For $L=12$ the necessary $\xi$ is already out of reach, explaining the effective N\'eel order we see in Fig.~\ref{fig:phasediagram}.
	Similarly, in the zigzag phase there is an extended region with a gapped symmetry-restored ground state. This is in keeping with 
	the sublattice transformation, which maps the zigzag to the N\'eel phase (in particular $\alpha = \frac{3}{4} \pi$ maps onto $\alpha = 0$).
	\begin{figure}[tb]
		\includegraphics[width=\linewidth]{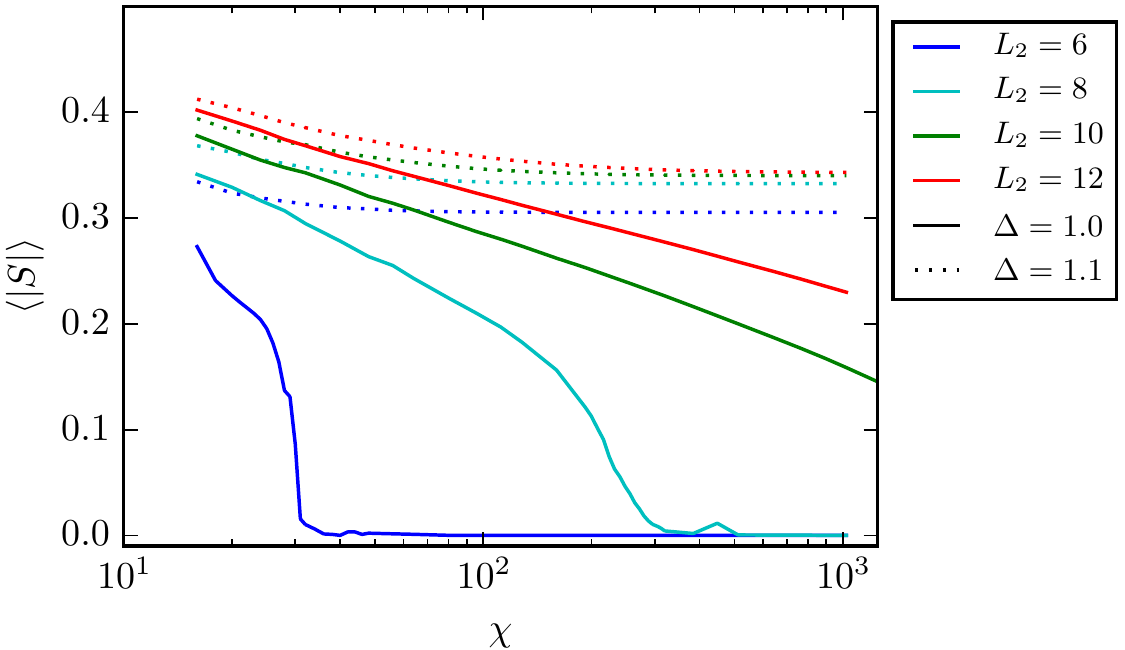}
		\caption{The absolute on-site magnetization for the pure Heisenberg AFM (solid) and for the AFM XXZ model with $\Delta = 1.1$ anisotropy (dashed) for different circumferences}
		\label{fig:RVB}
	\end{figure}

\section{Entanglement scaling of the gapless KSL}\label{app:entanglement_scaling}
	Matrix-product states (MPS) cannot capture algebraic ground state correlations. 
	However, increasing the bond dimension gives an increasingly accurate estimate of the  wave function. 
	\citet{calabrese2004} have shown that the entanglement entropy $S$ scales logarithmically with the correlation length $\xi$.
	In the MPS formalism, this is known as \textit{Finite-Entanglement Scaling} with $S_\chi = c/6 \log \xi_\chi $, where $\chi$ is the bond dimension of the MPS and $c$ is the \textit{chiral central charge} \cite{tagliacozzo2008,pollmann2009}. 

	\begin{figure}[tb]
		\includegraphics[width=\linewidth]{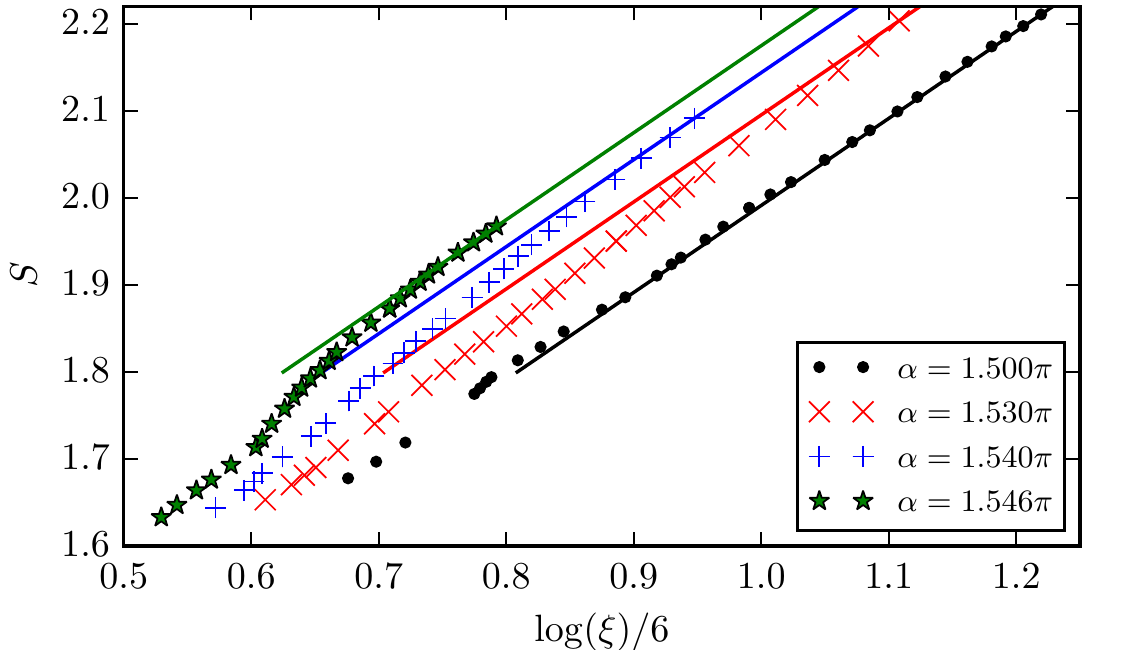}
		\caption{Entanglement entropy $S$ and logarithm of correlation length $\xi$ for different bond dimensions. The lines correspond to a central charge of $c=1$.}
		\label{fig:Slogxi}
	\end{figure}
	
	Fig.~\ref{fig:Slogxi} shows $S$ and $\log \xi$ for various MPS bond dimensions $\chi$ of up to $1024$. 
	The lines serve as a guide to the eye corresponding to a slope with $c=1$.
	We observe a  good match of the scaling for the pure Kitaev spin liquid  at $\alpha = 3/2 \pi$.
	This reflects the fact, that the KSL can be mapped to a free fermion problem with two Majorana cones in the first Brillouin zone, each contributing $1/2$ to the {central charge}.
	The gapless nature persists within the whole KSL phase and the scaling suggests $c=1$.

\section{Ground sectors of the KSL on the cylinder} \label{app:ksl_sectors}
	Similar to the plaquette operators $W_p = \prod_{j \in \hexagon} \sigma^{\gamma_j}_j$ we define a loop operator around the cylinder as
	\begin{equation}
		W_l = \prod_{j \in loop} \sigma_j^{\gamma_j}~,
	\end{equation}
	where $\gamma_i = \{x,yz\}$ corresponds to the bond that is not part of the loop at site $i$.
	Following \citet{kitaev2006}, $W_l$ can be expressed in terms of $\mathbb{Z}_2$ gauge field variables $u_{jk}$
	\begin{equation}
		\tilde W_l = \prod_{(j,k) \in loop} u_{jk}~.
	\end{equation}
	For our choice of  lattice periodicity, both loop operators are related by a minus sign.
	Thus, $\tilde W_l \rightarrow +1$ (periodic boundary condition of the fermions) translates to $W_l \rightarrow -1$, which corresponds to the gapless sector if the cylinder is chosen such that cuts in reciprocal space go through the nodes of the Majorana cones.
	The second sector  (antiperiodic boundary condition of the fermions) is always gapped and has a lower ground state energy than the gapless sector. 
	
	\begin{table}[tb]
		\begin{tabular}{l|c|c|cccc}
							&	ED 		& iPEPS		& DMRG			& 				& 					&	\\
			\hline
							&				&				& $L_2 = 6$ 	&				& $L_2 = 12$	& \\
							&				&				& gapped 		& gapless	& gapped 		& gapless \\
			\hline
			AF/KSL		&	0.488		&	0.487		& 0.484			& 0.494		& 0.485			& 0.487 \\
			KSL/ZZ		&	0.510		&	0.513		& 0.523			& 0.513 		& 0.514			& 0.512 \\
			%ZZ/FM		&	0.898		&				& 0.90			& 0.90		& 0.894			& 0.894	\\
			FM/KSL		&	1.399		&	1.432		& 1.405			& 1.44		& 1.421			& 1.428 \\
			KSL/ST 		&	1.577		&	1.557		& 1.573 			& 1.548		& 1.562			& 1.558 \\
			%ST/AF		&	1.812		&				& 1.81			& 1.81		& 1.815			& 1.815 \\
			\hline	
		\end{tabular}
		\caption{Transition points $\alpha/\pi$ for different circumferences sectors compared to exact diagonalization (ED)\cite{gotfryd2016} and infinite Projected Entangled Pair States (iPEPS)\cite{osorioiregui2014}.}
		% \matthias{Have to check values for gapped $L=12$ FM KSL phase again. Current values are obtained from very few points.}
		\label{tbl:pt}
	\end{table}
	
	Regarding the computation of the ground state, we can now make use of the loop operator and initialize DMRG with a state $|\psi\rangle$ that has $\langle \psi |W_l| \psi \rangle = \pm1$ depending on the desired sector. 
	Table \ref{tbl:pt} contains the phase transitions for the gapped and the gapless sector and compares it to {exact diagonalization} (ED) and {infinite Projected Entangled Pair States} (iPEPS).
	As the gapped sector has a lower energy, its stability is enhanced and widens the KSL phase.
	This effect is more pronounced for a small circumference $L_2 = 6$.
		
\section{Dynamics of $L_2=6$ cylinder} \label{app:L2_6_T40}
	
	In Fig.~\ref{fig:gamma} we show $\mathcal S(\boldsymbol k=0, \omega)$ for the same choices of $\alpha$ as in 
	\ifx\targetformat\undefined	% seperate main text / supplemental
		Fig.~5 in the main text,
	\else
		Fig.~\ref{fig:colorful},
	\fi	
	but now with a sharper $\omega$-resolution (corresponding to $T = 40$) which is possible due to a smaller circumference ($L_2 = 6$).
	The finer features are most likely discretization effects due to the finite circumference, but the main points are that the broadness in $\omega$-space persists despite a finer resolution, and that the high-energy feature gets squeezed downward as we get further away from the nearby spin liquid.
	Note that the latter is a meaningful statement and not just due to an overall $\alpha$-dependent scaling of the Hamiltonian since the minima of the spin bands (as shown in the first column of 
	\ifx\targetformat\undefined	% seperate main text / supplemental
		Fig.~5 in the main text)
	\else
		Fig.~\ref{fig:colorful})
	\fi
	do \emph{not} come down in energy (all at approximately $\omega = 0.4$).
	
	\begin{figure}[htb]
		\includegraphics[scale = .42]{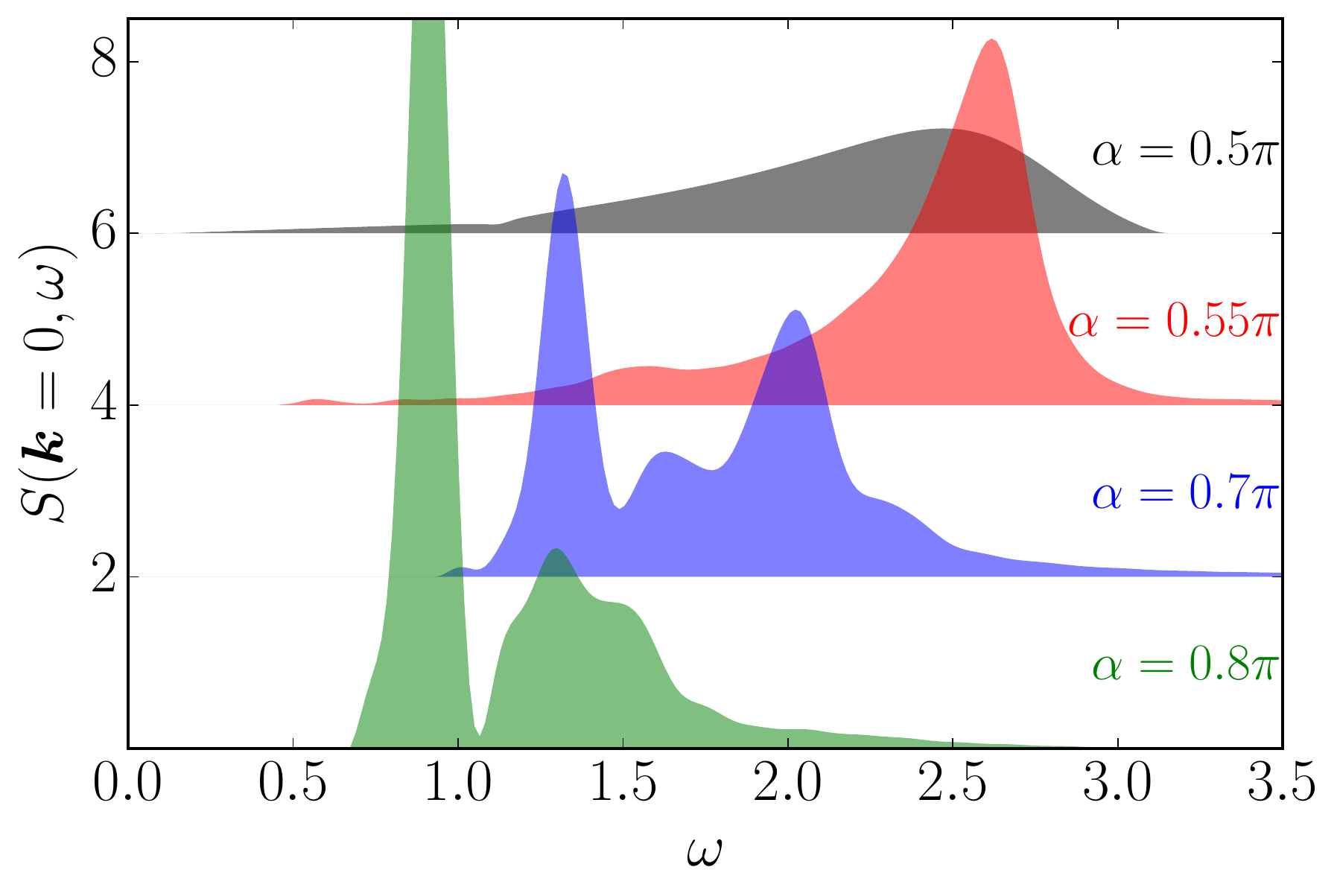}
		\caption{Complementing 
			\ifx\targetformat\undefined	% seperate main text / supplemental
				Fig.~5 of the main text:
			\else
				Fig.~\ref{fig:colorful}:
			\fi
			$\mathcal S(\boldsymbol k=0, \omega)$ for $\alpha = 0.5\pi$ (2D analytic result) and $\alpha = 0.55\pi,0.7\pi,0.8\pi$ (with $L_2 = 6$).}
		\label{fig:gamma}
	\end{figure}
	\fi
	
\end{document}